\documentclass[useAMS,usenatbib]{mn2e}
\usepackage[dvips]{graphicx}
\usepackage{color}

\title[Torus-fitting method for obtaining action variables]
{Torus-fitting method for obtaining action variables in two-dimensional 
Galactic potentials}
\author[H. Ueda,T. Hara,N. Gouda, T. Yano]
{H. Ueda$^{1}$\thanks{E-mail:email@address (HU); ueda@ipc.akita-u.ac.jp}
T. Hara$^{2}$\thanks{E-mail:email@address (TH); takuji.hara@nao.ac.jp} 
N. Gouda$^{2}$\thanks{E-mail:email@address (NG); naoteru.gouda@nao.ac.jp} 
T. Yano$^{2}$\thanks{E-mail:email@address (TY); yano.t@nao.ac.jp} 
\\
$^{1}$Faculty of Education and Human Studies, Akita University, Tegata-gakuen, Akita 010-8502, Japan\\
$^{2}$National Astronomical Observatory, Mitaka, Tokyo 181-8588, Japan\\
}

\begin{document}

\date{Accepted 2010 ???. Received 2010 ???; in original form 2010 July 25}

\pagerange{\pageref{firstpage}--\pageref{lastpage}} \pubyear{2010}

\maketitle

\label{firstpage}

\begin{abstract}
A phase-space distribution function of the steady state in 
galaxy models that admits regular orbits overall in the
phase-space can be represented by a function of three action 
variables. This type of distribution function in Galactic 
models is often constructed theoretically for comparison of the Galactic 
models with observational data as a test of the models. On the other hand, 
observations give Cartesian phase-space coordinates of stars. 
Therefore it is necessary to relate action variables and 
Cartesian coordinates in investigating whether the 
distribution function constructed in galaxy models can
explain observational data.
Generating functions are very useful in practice for this 
purpose, because calculations of relations between action variables 
and Cartesian coordinates by generating functions do not require 
a lot of computational time or computer memory in
comparison with direct numerical integration calculations of 
stellar orbits.
Here, we propose a new method called a torus-fitting method,
by which a generating function is derived numerically 
for models of the Galactic potential in which almost all orbits are regular.
We confirmed the torus-fitting method can be applied to major 
orbit families (box and loop orbits) in some
two-dimensional potentials.
Furthermore, the torus-fitting method is still applicable to 
resonant orbit families, besides major orbit families. Hence 
the torus-fitting method is useful for analyzing 
real Galactic systems in which a lot of resonant orbit families might exist.
\end{abstract}


\section{Introduction}

Our Galaxy is unique among galaxies in which we can observe detailed 
dynamics and kinematics of stars with high accuracy. 
These observations are performed by spectrometric observations 
and astrometric measurements.
We can obtain six phase-space coordinates for stars with astrometric 
measurements that provide
 five-dimensional phase-space coordinates
(three-dimensional positions and two-dimensional transversal velocities)
and also spectroscopic measurements that provide radial velocities.
Some modern space astrometry missions (Gaia 
\footnote{http://www.rssd.esa.int/index.php?project=GAIA\&page=index} 
and JASMINE \footnote{http://www.jasmine-galaxy.org/index.html}) 
will provide more than a thousand million accurate five-dimensional 
coordinates, making it possible to study the 
detailed and accurate current dynamical state of the Galaxy.

Construction of dynamical models of our Galaxy is currently 
anticipated, because we need models that can be compared 
with accurate observational data in the near future.
In this situation, we concentrate our attention on 
steady-state models of our Galaxy. 
This is because constructing steady-state models is not very 
difficult to accomplish, and is useful as a first step for 
investigation of the real Galactic structure.
In addition, steady-state models are known to be of 
fundamental importance even though the Galaxy cannot be 
in a steady state (Binney 2002).
In any case, here, we devote attention to steady-state models 
of our Galaxy. 

Phase-space distribution functions for all matters in the Galaxy 
are fundamental for describing the dynamical structure of our Galaxy.
In general, a phase-space distribution function is a 
seven-dimensional function $f({\bf x},{\bf p},t)$ of three-dimensional positions, 
three-dimensional velocities and (one-dimensional) time.  
If we suppose that the dynamical state of the Galaxy is steady,
this function is expressed as a six-dimensional
function $f({\bf x},{\bf p})$.
However, treatment of a six-dimensional function is still
very complicated.
Fortunately, the strong Jeans theorem suggests that a distribution 
function of a steady state model in which almost all 
orbits are regular 
with incommensurable frequencies, may be
presumed to be a function of three independent isolating 
integrals (Binney \& Tremaine 1987).
Any three-dimensional orbit that admits three
isolating integrals forms a three-dimensional torus in phase space (Arnold 1989).
This suggests that a phase-space distribution function of the
steady-state model may be 
expressed as $f(J_1,J_2,J_3)$, where $J_i$ are action variables
(isolating integrals). 

On the other hand, 
observations do not provide action-angle coordinates but 
Cartesian coordinates of stars. So it is necessary to 
relate action variables $({\bf J})$ to Cartesian coordinates 
$({\bf x},{\bf p})$ to compare theoretical 
models with observational data.
The relationships are estimated in principle 
by direct numerical integration 
of orbits of stars, 
$ J_i = \frac{1}{2 \pi} \oint_{\gamma_i} {p} d {q}$.
However this direct integration method is not practical 
for application to the real 
observational data that will be provided in near future.
For example, Gaia will bring us information concerning the positions 
and velocities of one billion stars. 
This means that it is necessary to estimate 
$({\bf x},{\bf p})  \Leftrightarrow ({\bf J})$ for one billion cases.
In addition, we have to examine many Galactic potential models that 
include many free parameters. It is apparent that the number of 
observed Cartesian coordinates of stars times the number of models 
is a terribly large number. We need this large number of relations 
between the Cartesian coordinates and action variables.
Hence this method by the direct numerical integration of orbits 
requires a lot of computational time and furthermore vast memories 
in computers. It is not practicable to use this method in 
real applications.

McGill \& Binney (1990) proposed that the use
of a generating function that relates
$({\bf x},{\bf p})$ to $({\bf J})$ has a significant advantage
over the direct numerical integration of orbits, because the
use of the generating function can 
reduce computational time 
and the amount of memory required for  computers. This is because 
a generating function can be expressed by a Fourier expansion with 
the relatively small number of Fourier coefficients. 
Furthermore, if generating functions can be derived at some 
values of action variables, then the generating functions at 
other values of the action variables can be easily derived using an interpolation technique as shown in 3.1 and 3.2.
As just described, the use of a generating function is a useful 
and practical method for relating action variables to Cartesian 
coordinates.

Moreover, McGill \& Binney (1990) suggested a method for making a 
generating function using an iterative approach.
This is called a torus construction method, which is briefly 
reviewed in Section 2.
This method is very elegant, and has been suggested as being 
applicable to some Galactic 
models that have two-dimensional gravitational potentials.
However, we find that this method has some weaknesses in terms of practical use 
in some cases.
For example, the torus construction method requires a complicated 
process using a perturbation method to reconstruct tori of resonant orbits.
Kaasalainen (1994) developed a method for perturbative calculations for
reconstructing the resonant tori. In this perturbation method,
we should take into consideration higher-order terms that can be 
ignored in normal perturbation methods.
This is therefore very complicated when one applies this method to real systems
with a lot of resonant orbits.

In this paper, we propose a new approach, a torus-fitting
method, which makes generating functions 
based on numerical integration of only some typical orbits.
Our torus-fitting method has an advantage over 
the torus construction method in application to tori of resonant orbits. 
Treatment of resonant orbits is important for applying to some 
gravitational potentials in some galaxy models such as asymmetric 
potentials and also the real 
Galactic potential.
The torus-fitting method is practical and very useful for making 
generating functions for any tori. We demonstrate the usefulness 
of the torus-fitting method by applying this method to some 
two-dimensional galactic potentials, some of which provide resonant 
orbits.

As mentioned above, the purpose of this paper is to estimate 
generating functions using the torus-fitting method, and to 
evaluate the relation between $({\bf J})$ and $({\bf x},{\bf p})$. To obtain the restriction of a theoretical model for comparing 
the phase-space distribution function of a Galaxy model and 
observational data, it is required and important to obtain the 
relation between $({\bf J})$ and $({\bf x},{\bf p})$. As stated, the phase-space 
distribution function is a function of $({\bf J})$ with numbers equal to 
the space dimension of a system under a condition that almost 
all orbits of stars are regular. On the other hand, observational 
data is given as $({\bf x},{\bf p})$, so the relation between $({\bf J})$ 
and $({\bf x},{\bf p})$ is needed. 
Torus fitting is a practical method for relating action variables to 
Cartesian coordinates, and it is therefore important. 

Note that the torus-fitting method can be applicable only when almost 
all orbits are regular or can be regarded as approximately regular. 
As this method cannot be applicable under a situation that chaotic 
orbits are dominant, we do not treat this case. Because the torus-fitting method cannot handle the chaotic orbits, one may think 
that this is not practicable. However application to some actual 
systems can be possible, and we discuss this in section 4. Even if 
we only consider the Galaxy model that almost all orbits are regular, the structure of torus on phase-space is in general 
complicated, i.e., resonance orbits appear in addition to major box 
and loop orbits. The advantage of the torus-fitting method is that 
it can be applicable to complicated torus structures containing 
resonance torus, and we discuss this in subsection 4.1.

The remainder of this paper is organized as follows: an
overview of the torus construction method is described in Section 2.
Explanation of our torus-fitting method and application to the major
families of the orbits are given in Section 3.
Application of our torus-fitting method to resonant orbits
is given in Section 4.
Finally, we provide a discussion in Section 5.

\section{Torus construction method}

The torus construction method was developed by McGill \& Binney (1990),
Binney \& Kumar (1993), Kaasalainen \& Binney (1994),
and Kaasalainen (1994,1995),
and we briefly review their method for constructing tori in general 
gravitational potentials.
The action-angle variables are extremely useful if the coordinate 
transformation 
$({\bf x},{\bf p})  \Leftrightarrow ({\bf J})$
can be performed analytically. 
The analytic transformation can be done only in Hamiltonian systems 
for harmonic oscillators and isochrone (generalizations of the 
Kepler potential). We call these Hamiltonians ``toy'' Hamiltonians 
hereafter.
On the other hand, an analytical expression of the action-angle variables 
of a Hamiltonian system with a general gravitational potential cannot be obtained.
Here we refer to these general Hamiltonians of systems for which 
we want to get the relations between the action-variables and 
Cartesian coordinates as ``target'' Hamiltonians.
If almost all orbits in the target 
$N$-dimensional Hamiltonian system are regular,
this system has $N$ isolating integrals ($N$ action variables). 
Hence the orbits form $N$-dimensional tori in the phase space.
We refer to these tori obtained from toy and target Hamiltonians as toy and
target tori, respectively.
McGill \& Binney (1990) 
obtained relationships between the action variables in the toy 
Hamiltonian and those in the target Hamiltonian by an iterative 
approach shown below, without the direct numerical integration of 
the trajectories (orbits) on the target tori.

We show the torus construction method as follows:
Let $H_0$ be a toy Hamiltonian, and $({\bf J},{\bf \theta})$
the action-angle coordinates of $H_0$.
On the other hand, $H$ represents a target Hamiltonian with the 
action-angle variables $({\bf J'},{\bf \theta'})$. Note that 
$({\bf J},{\bf \theta})$ are analytically expressed as a function of the 
Cartesian coordinates. Relationships between $({\bf J},{\bf \theta})$ 
and $({\bf J'},{\bf \theta'})$ are determined by a generating 
function $S({\bf \theta'},{\bf J'})$. For a generating function of the $F_2$-type
(Goldstein et al. 2002), we have
\begin{equation}
{\bf J}({\bf \theta},{\bf J}') = {\partial S\over\partial{\bf \theta}}, 
\ \ \ \
{\bf \theta}'({\bf \theta},{\bf J}') = {\partial S\over\partial {\bf J}'}.
\end{equation}
Geometrically, the generating function maps the toy tori into the 
target tori.
As is well known, this function is expanded when the system
has a periodic condition. In this case, we find that
\begin{equation}
S({\bf \theta},{\bf J}') = {\bf \theta} \cdot {\bf J}' - i\sum_{{n}\ne 0}
S_{n}({\bf J}')e^{i{n}\cdot {\bf \theta}},
\end{equation}
where the first term is the identity transformation.
From equations (1) and (2), we obtain the following relation,
\begin{equation}
{\bf J} = {\bf J}' + \sum_{{\bf n}\ne 0} {\bf n} 
S_{\bf n} e^{i{\bf n}\cdot{\bf \theta}}.
\end{equation}
As the coefficients of the generating function $S_{\bf n}$ are real and 
$S_{\bf -n}=-S_{\bf n}$, the above relation is modified as
\begin{equation}
{\bf J} = {\bf J}' + 2 \sum_{{\bf n} > 0} {\bf n} 
S_{\bf n} \cos({\bf n}\cdot{\bf \theta}).
\end{equation}
If we can get the correct $S_{n}$, the action variable of the target 
Hamiltonian ${\bf J'}$ can be
expressed as a function of Cartesian coordinates 
${\bf J'}={\bf J'}({\bf x},{\bf p})$
through the action variables of the toy Hamiltonian analytically expressed as a function of the Cartesian coordinates.
In this way, the main objective of the torus construction method is
to derive $S_{n}$ for target Hamiltonians of galaxy models. 

How can we determine $S_{n}$ numerically? 
McGill \& Binney (1990) choose $N_p$ points on the target 
torus, and consider the variance of total energies of these points.
The variance must be zero, because the total energy at each point has to 
be constant (note that the total energy in 
the system we consider here is conserved). 
If it is not zero, this means that the transformation 
$({\bf x},{\bf p})  \Leftrightarrow ({\bf J'})$
is not performed correctly. In other words, one fails to determine
$S_{n}$ properly. Beginning from tentative values of 
$S_{n}$ (initial and trial value), we reduce the variance close to zero
by changing $S_{n}$ properly.
The outline of the torus construction method is as follows:
\begin{enumerate}
  \item Choose $N_p$ sets $({\bf J'},{\bf \theta_i})$  
on a torus with constant ${\bf J'}$.
  \item Set trial generating function coefficients $S_{n}$.
  \item Transform $({\bf J'},{\bf \theta_i})$ to $({\bf J_i},{\bf \theta_i})$ using $S_{n}$.
  \item Calculate $({\bf x_i},{\bf p_i})$ from $({\bf J_i},{\bf \theta_i})$. 
  \item Estimate $H_i$ and $\chi^2 = \frac{1}{N_p} 
\sum_{i=1}^{N_p} \vert H_i - \bar H \vert^2$.
  \item Iterate (ii) $\sim$ (v) to minimize $\chi^2$, and finally we get \\
$~~~~~~~~~~$ $S_{n}$.
\end{enumerate}
Note that $N_p$ sets of $({\bf J'},{\bf \theta})$ are chosen
under a condition that 
${\bf J'}$ is constant, and $\bar H $ is defined as 
$\bar H = \frac{1}{N_p} \sum_{i=1}^{N_p}  H_i$.
Note that this method does not use the direct numerical
integration of the trajectories on the target tori.

Although the torus construction algorithm is clear, some
difficulties exist with this method.
First, we must prepare a toy torus before constructing a
target torus. It is well known that major orbits are classified into 
two families, i.e. the box orbit family and the loop orbit family.
The toy Hamiltonian must be set as a harmonic oscillator when an 
orbit of the target Hamiltonian is the box type, and as an isochrone 
when an orbit of the target Hamiltonian is the loop type.
As Kaasalainen \& Binney (1994) noted, a successful torus 
construction method depends strongly on the choice of the toy Hamiltonian, 
and this is an essential part for bringing a successful conclusion 
in this method.
A toy Hamiltonian should be prepared
without any 
direct numerical integration of trajectories on the target tori 
if only the above iteration method is used.
Although we may determine the toy Hamiltonian by trial and error 
in the iterative approach by changing the toy Hamiltonian, 
this procedure makes the method complicated.

Next, we must determine several hundred coefficients of the 
generating function based only on the condition by which
the variance of total energies at each point $N_p$ should be minimized.
In this method, there is no guarantee
that the iteration of algorithm converges to real generating functions.
That is, we do not confirm
whether the torus obtained by the torus construction method
corresponds to the target torus.
It is therefore very difficult to construct the torus without
derivation of target tori by numerical integration of orbits.

Finally, the torus construction method requires a 
very complicated procedure when applied
to general gravitational potentials that provide resonant orbits. 
If we wish to deal with a resonant orbit or resonant torus, we have to 
combine the torus construction method 
with perturbation (Kaasalainen 1994).
Treatment of the resonant orbit is important when we 
investigate many target Hamiltonians with general gravitational 
potentials and also the real Galactic system.
Thus the torus construction method is not necessarily practical for many target 
Hamiltonians with resonant tori.

\section{Torus-fitting method}

\subsection{Procedure in the method}

We propose a new method for making generating functions, which is 
practical for many target Hamiltonians with resonant tori. In this method, 
we use direct calculations of some tori in a target Hamiltonian by 
numerical integration of orbits. 
We do not need direct calculations of all tori, but only some typical 
ones and we can estimate generating functions for other tori with 
the results of the direct calculations of some tori.
The outline of a new method, the torus-fitting method, is as follows:
\begin{enumerate}
  \item Set an initial phase space position $({\bf x},{\bf p})$ of a test \\
$~~~~~~~~~~$ particle in a given potential in a target Hamiltonian \\
$~~~~~~~~~~$ system.
  \item 
Follow numerically a trajectory (orbit) of the test \\
$~~~~~~~~~~$ particle under the given potential, and create a  \\ 
$~~~~~~~~~~$ target torus, which can be represented on the surface\\
$~~~~~~~~~~$ of section (Poincare section). In addition, store some \\
$~~~~~~~~~~$ phase space positions $({\bf x},{\bf p})$ on the trajectory (orbit). \\
  \item Estimate action variables ${\bf J'}
 = \frac{1}{2 \pi} \oint {p} d {q}$ of this test \\
$~~~~~~~~~~$ particle.
  \item Determine the appropriate type of a toy Hamiltonian \\
$~~~~~~~~~~$ according to the shape of the orbit (the torus) shown\\
$~~~~~~~~~~$ on the surface of section, that is, the type of the\\
$~~~~~~~~~~$ orbit (box or loop). The Hamiltonian of the harmonic\\
$~~~~~~~~~~$ oscillator is adopted as a toy Hamiltonian for the \\
$~~~~~~~~~~$ box-type orbits, and the Hamiltonian of the isochrones\\
$~~~~~~~~~~$ is adopted as the toy Hamiltonian for the loop-type\\
$~~~~~~~~~~$ orbit. Fix the free parameters included in \\
$~~~~~~~~~~$ the toy Hamiltonian as first trial values. These values \\
$~~~~~~~~~~$ can be determined under the condition that the  sha- \\
$~~~~~~~~~~$ pe of the toy torus on the surface of section corres-\\
$~~~~~~~~~~$ ponds to that of the target tori as closely as possible. \\
$~~~~~~~~~~$ Refer to 3.2 for details. On the other hand, equation \\
$~~~~~~~~~~$ (4) shows the average value of ${\bf \overline{J}} \equiv \int {\bf J} d {\bf \theta }$ should be \\
$~~~~~~~~~~$ equal to ${\bf J'}$. If this condition is satisfied with less \\
$~~~~~~~~~~$ than several percent errors, the trial values of the \\
$~~~~~~~~~~$ free parameters are adopted as final values in the \\
$~~~~~~~~~~$ toy Hamiltonian. If not, the values of the free param- \\
$~~~~~~~~~~$ eters will be changed by trial and error until the \\
$~~~~~~~~~~$ condition is satisfied with less than several percent\\
$~~~~~~~~~~$ errors. Note that when the values of\\
$~~~~~~~~~~$ free parameters in the toy Hamiltonian are fixed as\\
$~~~~~~~~~~$ the final ones, we can use the same values for other\\
$~~~~~~~~~~$ test particles if the type of tori for other test particles\\
$~~~~~~~~~~$ is the same. This fact will be shown clearly in\\
$~~~~~~~~~~$ application of this method as explained in 3.2.
\item  Translate analytically phase space positions $({\bf x},{\bf p})$\\
$~~~~~~~~~~$ of the test particles obtained from the direct \\
$~~~~~~~~~~$ orbit integration into the action and angle valuables\\
$~~~~~~~~~~$ $({\bf J},{\bf \theta})$ of the toy Hamiltonian with the fixes values\\
$~~~~~~~~~~$ of parameter.Then, the generating  function coefficients\\
$~~~~~~~~~~$  $S_{n}$ are determined by the least-squares method from\\
$~~~~~~~~~~$  equation (4). See 3.2 for details.
\end{enumerate}
In this way, we get the generating function at a particular 
value of ${\bf J'}$ associated with the test particle.
For some other values of ${\bf J'}$ associated with some other 
test particles, the same procedure shown above allows derivation of 
generating functions at some other values of ${\bf J'}$. 
Sandres \& Binney (2014) suggested a similar method to obtain ${\bf J'}$ and $S_{\bf n}$ for numerically integrated orbits. In deriving  the generating function coefficients $S_{\bf n}$, they also derive ${\bf J'}$ simultaneously without the procedure (iii) by the least-squares method. The difference between their procedure and ours is not important, and the point we would like to note is the following:
They performed the procedure on each particle to obtain ${\bf J'}$ and $S_{\bf n}$.  However, we do not repeat the same procedure to 
get generating functions at all other values of ${\bf J'}$.  
As shown in 3.2 (see Fig.4 and Fig.5) and in 3.3 (see Fig.7), coefficients of generating 
functions are smooth function of J' if the torus
is the same type. Hence we can get coefficients of generating functions at 
any values of J' by interpolating some typical coefficients. This fact reduces the computational time and amount 
of computer memory. It should be remarked that this method is still applicable to making 
generating functions for resonant tori, although we need additional techniques 
shown in section 4.
Details of the torus-fitting method are explained 
in the next subsection by showing the application of this method to 
some galaxy models.

\subsection{Application to logarithmic potential}

In this subsection, we demonstrate that the torus-fitting method
works well for two-dimensional Galactic potentials.
As a first example, we show the case of the two-dimensional 
logarithmic potential,
\begin{equation}
\Phi = \frac{1}{2} \ln (x^2 + \frac{y^2}{q^2} + R_c^2),
\end{equation}
where $q$ and $R_c$ are constants. 
As the logarithmic potential with $q=0.9$, $R_c=0.14$ was examined 
by Binney \& Tremaine (1987), we use these 
values in the following discussion.
Shapes of orbits with total energy $E \cong -0.337$ are displayed in
Fig.3-7 in Binney \& Tremaine (1987).

According to the procedures (i) and (ii) shown in 3.1, we set some test
particles and construct invariant tori (i.e. target tori) by numerical 
integration of the orbits of the test particles. 
Fig.1 shows target tori for three test particles on the 
surface of section with $y=0$. 
As the values of the total energy are $H \cong -0.337$ for the 
test particles,
this figure is the same as the Fig.3-8 in Binney \& Tremaine (1987).
In addition, some phase space positions $({\bf x},{\bf p})$ are stored
to follow the trajectories of the test particles. 
About 500000 points are stored in each torus of each test particle.
\begin{figure}
 \begin{center}
  \includegraphics[width=87mm]{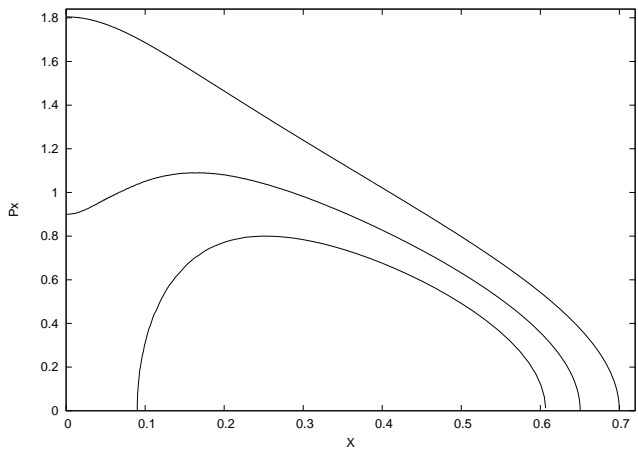}
 \end{center}
\caption{The surface of section $(x,p_x)$ with $y=0$ in the 
logarithmic potential with $q=0.9$ and $R_c=0.14$. The target tori 
derived by the numerical calculations of the orbits of three test 
particles are shown on the surface of section.
The particles on these tori have the same 
energy $E \cong -0.337$. The outermost curve
corresponds to the torus with action variables 
$(J_1', J_2') \cong (0.48,0.008)$, the middle
curve corresponds to the torus with $(J_1', J_2') \cong (0.34,0.12)$
and the innermost curve corresponds to the torus with 
$(J_1', J_2') \cong (0.2,0.24)$.
}
\end{figure}

Next, we estimate the action variables $J_i'$ according to the 
procedure (iii).
We cannot obtain analytically action variables for the logarithmic potential.
However, values of the action variables $J_i'$ of a torus can 
be derived as follows;
\begin{equation}
 J_i' = \frac{1}{2 \pi} \oint_{\gamma_i} {\bf p} d {\bf q},
\end{equation}
where ${\bf q}$ and ${\bf p}$ are the generalized coordinate 
and generalized momentum, 
and $\gamma_i$ is a basis for the one-dimensional cycle on the torus.
The values of action variables are estimated by carrying out orbital integration 
of test particles and construct invariant tori. In general, a two-dimensional 
torus has two independent bases for the one-dimensional cycles 
$\gamma_1, \gamma_2$, and each define action variables from a formula (6). 
It is well known that values of action variables do not depend on the shape 
of $\gamma$ (Arnold 1989),
and here, we choose $\gamma$ with constant angle and radius.
To calculate action variables definitely, it is necessary to pursue 
integration of test particles until values of action variables
converge. In this paper, a sufficient number of points $({\bf x},{\bf p})$ are 
used for deciding the value of action variables.
In Fig.1, the outermost curve
corresponds to the torus whose values of the action variables 
$(J_1', J_2') \cong (0.48,0.008)$, the middle curve corresponds to the 
torus with $(J_1', J_2') \cong (0.34,0.12)$
and the innermost curve corresponds to the torus with 
$(J_1', J_2') \cong (0.2,0.24)$.

Next we determine a toy Hamiltonian according to the procedure (iv). 
As stated, two candidates of the type for a toy Hamiltonian
exist, i.e., the harmonic oscillator type and isochrone potential type. 
The type of a toy Hamiltonian should be the same as one of a target torus.
The surface of section in Fig.1 tells us clearly how to choose the toy 
Hamiltonian. The harmonic oscillator type should be adopted
for the tori with $(J_1', J_2') \cong (0.48,0.008)$ and $(0.34,0.12)$ 
because these tori represent box orbits.
On the other hand, the isochrone type should be adopted for the torus with
$(J_1', J_2') \cong (0.2,0.24)$, because this torus represents a loop orbit.
A toy Hamiltonian for the harmonic oscillator 
type is given by the Cartesian coordinates as follows, 
\begin{equation}
H_H(x,y,p_x,p_y)=\frac{1}{2} (p_x^2 + p_y^2 + 
\omega_x x^2 + \omega_y y^2),
\end{equation}
where $\omega_x$ and $\omega_y$ are free parameters that must be determined
according to the procedure (iv).
On the other hand, a toy Hamiltonian for the isochrone potential type, $H_I$
is represented by the plane polar coordinate as follows; 
\begin{equation}
H_I(r, \phi, p_r, p_{\phi}) = \frac{1}{2} p_r^2 +
\frac{p_{\phi}^2}{2r^2} - \frac{k}{b+\sqrt{b^2+r^2}},
\end{equation}
where $k$ and $b$ are free parameters.

Next we will determine appropriate values of the free parameters 
included in the above toy Hamiltonians. As mentioned in 3.1, these 
values can be determined under the condition that the shape of the 
toy torus corresponds to that of the target tori as closely as possible. 
As a first example, we consider the case that the toy Hamiltonian is 
the harmonic oscillator type. As is well known,   
the shape of the target torus for the box orbit on the surface of 
the section is changed according to the value of $\omega_x$. 
When $\omega_x$ is small, the shape of the torus is horizontally long, 
but the shape is changed to be vertically long when $\omega_x$ is 
made large.
Using this fact, we can adjust $\omega_x$ so as that the shape of 
the toy torus is similar to that of the target torus.
\begin{figure}
 \begin{center}
  \includegraphics[width=60mm,angle=-90]{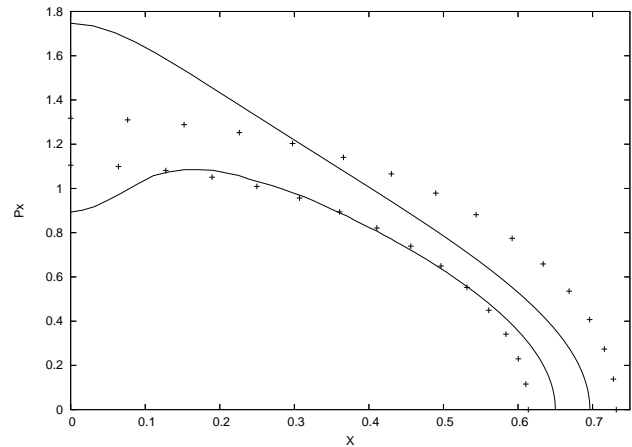}
 \end{center}
\caption{The same surface of section as shown in Fig.1, but the target 
tori for the box-type orbits are shown by the solid curves. 
Two solid curves represent the tori
with $(J_1', J_2') \cong (0.48,0.008)$
and $(J_1', J_2') \cong (0.34,0.12)$ (these are the same as shown in 
Fig.1).
The plus symbols show the toy torus with $\omega_x=1.8$.
}
\end{figure}
By this adjustment of $\omega_x$, we found that the shape of the toy 
torus is similar to the target torus when $\omega_x =1.8$. Fig.2 
shows the comparison of the toy torus with the adjusted value of the free parameter ($\omega_x =1.8$), with the target torus. We finally set $\omega_x =1.8$. 
Furthermore we should determine the value of $\omega_y$. 
The same procedure leads to determination of the value of $\omega_y$.
Here, however, as a trial,
$\omega_y$ is set so as the average of the total toy Hamiltonian energy of $N_p$ points 
becomes $E \sim -0.337$ (the same energy as the total energy of a 
test particle on the target torus). 
In this trial, 
the values of $(\omega_x$, $\omega_y)$ are $(1.8,3.5)$.
Therefore, the iteration process is not needed in this case.
The (shape of) target torus derived by the direct numerical 
calculation brings us necessary information for determining the 
appropriate values of the free parameters with good accuracies.
 
It should be noted that when the values of free parameters in the 
toy Hamiltonian are fixed as the final ones, we can use the same 
values for other test particles if the type of tori for the 
other test particles is the same.
Namely, the generating functions at different values of J' can 
be determined by one set of the values of the free parameters 
determined only at a particular value of J'. This fact is proved to 
be correct by applying this method to some potential models. 
Fig.3 shows the target tori reconstructed using the same set of values of the free parameters even at different J' corresponds very 
well to the tori derived by the direct numerical calculations. 

In the case of the loop orbit, we focus on the case with 
${{\bf J'_1} \cong 0}$ when we determine the values of the free 
parameters in the isochrones potential.
This is because this case is a very particular kind of 
invariant torus, because the volume of invariant torus 
as well as all coefficients of the generating function nearly equal zero, and the free parameters strongly influence the shape
of the invariant torus.
By the adjustment of the values of the free parameters, we find 
that $k=1.$ and $b=0.14$ are appropriate sets of the values that the shape of the toy torus becomes nearest to the shape of the target torus. We confirmed that the trial values of the free parameters are good 
enough to satisfy the average condition of equation (4) mentioned before 
within a few percent. Therefore we do not need the iteration process 
also in the case of the loop orbit.

Next, shown in the procedure (v), as the values of the free 
parameters in the toy Hamiltonian is determined, we can analytically 
translate the phase space positions on a target torus $({\bf x_j},{\bf p_j})$ 
of a test particle into the action and angle variables 
$({\bf J_j},{\bf \theta_j})$ $(j=1,2,\cdots)$ of the toy Hamiltonian.
Now we should prepare representative $N_p$ sets 
$({\bf J_i},{\bf \theta_i})$ $(j=1,2,\cdots ,N_p)$ to derive the 
coefficients of the generating function, $S_{\bf n}(J')$, by using equation (4).
It is necessary to get $N_p$ sets that include almost all the whole range 
of the values of $({\bf J_i},{\bf \theta_i})$ uniformly to derive
$S_{\bf n}(J')$ accurately. However, the set of $({\bf J_i},{\bf \theta_i})$ translated analytically from the stored data of the Cartesian coordinates $({\bf x_i},{\bf p_i})$ of the orbit of the test particle are not completely uniform.
Because distribution of $({\bf J_i},{\bf \theta_i})$ depends on the target Hamiltonian and the values of free parameters of the toy Hamiltonian, it is not appropriate to evaluate the generating function coefficients in (4) using the inverse Fourier transformation. Therefore, we derive the coefficients of the generating function $S_{\bf n}$ from the least-squares method.

Equation (4) in the two-dimensional case is expressed as
\begin{eqnarray}
J_1 & = & J_1' + 2 \sum_{\bf n > 0} n_1 
S_{n_1 n_2} \cos(n_1 \theta_1 + n_2 \theta_2), \\
J_2 & = & J_2' + 2 \sum_{\bf n > 0} n_2 
S_{n_1 n_2} \cos(n_1 \theta_1 + n_2 \theta_2),
\end{eqnarray}
where ${\bf n}=(n_1,n_2)$.
We found that $0 \le |n_1|, |n_2| \le 16 \sim 20$ are necessary and 
sufficient to reconstruct the target tori accurately.Here we adopt $n_1=0,\cdots ,18,~n_2=-18,\cdots ,18$; 
the total of 702 coefficients are needed to reconstruct the target torus 
(note that $S_{0,0} = 0$). But the necessary number of coefficients depends on the values of free parameters.
So if we choose appropriate values of free parameters, we can reduce the number of coefficients.
In this way, we get the generating function at the particular value of 
${\bf J'}$ for the test particle.
By using the generating function, we can determine the value of the 
toy Hamiltonian action variable ${\bf J}$ for any value of the angle 
variables ${\bf \theta}$.

Fig.3 shows the same surface of section as shown in Fig.1.
In Fig.3, the solid curves represent the target tori derived by the direct
calculation of the orbits of
three test particles. The plus symbols show the reconstructed target 
tori by use of the torus-fitting method. We find that the reconstructed 
tori correspond very well
to the (true) tori derived directly from the orbits. 
This fact proves that the torus-fitting method works very well for
both box and loop orbits (major families of the orbits).

\begin{figure}
 \begin{center}
  \includegraphics[width=85mm]{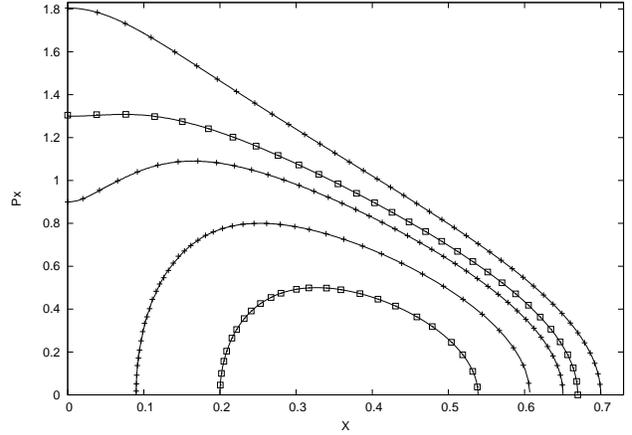}
 \end{center}
\caption{The same surface of section as shown in Fig.1. The solid 
curves represent the target tori derived by the numerical calculation 
of the orbits of five test particles. The plus symbols show the 
reconstructed target tori by use of the torus-fitting method. 
 See the text for details. The square symbols represent the target tori derived by the tours-fitting method in which the generating functions are estimated by the interpolation technique. See the text for details.} 
\end{figure}
Furthermore we show 
that coefficients of generating functions of 
J' are smooth functions for any type of orbit. 
This character of the generating function is 
very important in the torus-fitting method as mentioned below.
Fig.4 shows the coefficients of the generating function
as functions of ${J'_1}$ for the box-type orbit.
The six coefficients taken in order of descending amplitudes of $S_{\bf n}$ are shown and they are calculated at
$J_1' \cong 0.34,0.39,0.45,0.48,0.49$ by the 
procedures (i) $\sim$ (v) mentioned in 3.1.
The plus symbols represent the amplitudes of the coefficients at 
these values of ${\bf J'}$. Furthermore, in Fig.4, each solid line 
shows the linear interpolation line connected with the plus symbols 
for each coefficient. We can see that the interpolation lines are 
very smooth functions of ${\bf J'}$ and so the interpolation 
technique for getting the values of the coefficient for any value of 
${\bf J'}$ can be used. 
We calculate the values of the coefficients derived by the 
procedure (i) $\sim$ (v) at $J_1' \cong 0.37,0.42,0.46$. 
These values are shown in Fig.4 by the crosses.
We can see that these marks correspond very well to the interpolation 
lines and so we confirm that the linear interpolation technique works 
very well.
Moreover we confirmed that other coefficients of the generating 
function (not shown in Fig.4) are also smooth functions of 
${\bf J'}$ and the interpolation 
technique can be used for other coefficients.
This fact is confirmed for the loop-type orbit.
We also examined some other cases of different
potentials and this fact can be applied to the other cases (see 3.3).
\begin{figure}
 \begin{center}
  \includegraphics[width=60mm,angle=-90]{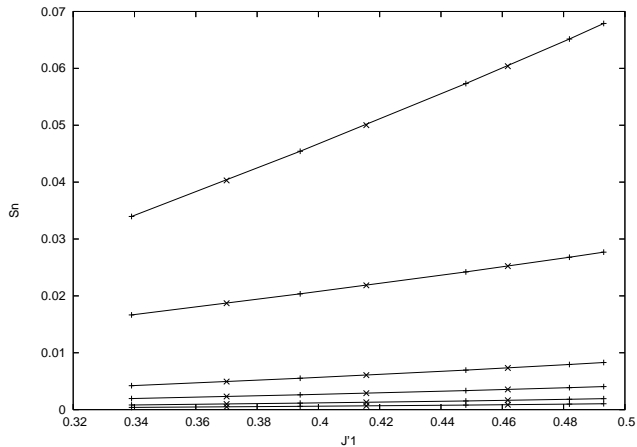}
 \end{center}
\caption{
The coefficients of the generating function as functions of ${J'_1}$ 
for the box-type orbit. The six coefficients taken in 
order of descending amplitudes of $S_{\bf n}$ are shown as 
functions of ${J'_1}$ for the box-type orbit. 
They are calculated by the procedure (i) $\sim$ (v) at
$J_1'\cong 0.34,0.39,0.45,0.48,0.49$.
The plus symbols represent the amplitudes of the coefficients at these 
values of ${\bf J'}$. Each solid line shows the linear interpolation 
line connected with the plus symbols for each coefficient. 
Furthermore the values of the coefficients derived by the procedure (i)
$\sim$ (v) at $J_1' \cong 0.37,0.42,0.46$ are shown by the crosses.
}
\end{figure}
\begin{figure}
 \begin{center}
  \includegraphics[width=60mm,angle=-90]{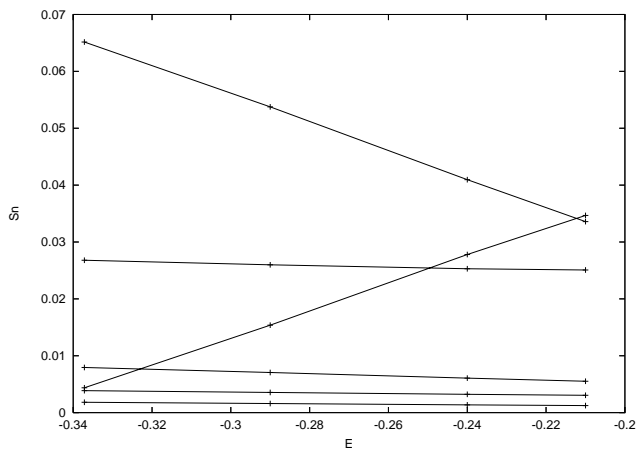}
 \end{center}
\caption{
The coefficients of the generating function as functions of ${E}$ 
for the box-type orbit. The six coefficients taken in 
order of descending amplitudes of $S_{\bf n}$ are shown as 
functions of ${E}$ for the box-type orbit. 
They are calculated by the procedure (i) $\sim$ (v) at
$E \cong -0.34,-0.29,-0.24,-0.21$.
The plus symbols represent the amplitudes of the coefficients at these 
values of ${E}$. Each solid line shows the linear interpolation 
line connected with the plus symbols for each coefficient. 
}
\end{figure}

Here, we mention some comments about Fig.4. Although the generating 
function is a function of two variables $ S_n = S_n (J_1', J_2')$, 
$S_n$ in Fig.4 is expresses as a function of one variable $J_1'$. 
This is because here we draw a Poincare section (Fig.1 or Fig.3) 
under the condition that a total energy is constant
($E \cong -0.337$). As the total 
energy of a system is a function of $J_1'$ and $J_2'$, i.e. 
$E = E (J_1', J_2')$, this means that $J_2'$ is automatically 
determined if one set $J_1'$. 
Because a generating function is originally a function of two 
variables, it is necessary to confirm the behavior of the generating 
function as a function of $E$ under the condition that the 
$J_1'$ is constant ($J_1' \cong 0.48$). The result is shown in Fig.5, and we 
also confirm that the generating function changes smoothly as a 
function of $E$. This means that the interpolation works well, 
and the torus fitting is a practical method to construct torus structures.

To confirm this fact, we reconstruct some tori using the interpolation technique.
For example,
we estimate the generating function at $J_1' \cong 0.43$ from the values of the 
coefficients of generating functions at 
$J_1' \cong 0.4$ and $J_1' \cong 0.49$, and reconstruct the target torus that represents the square symbols in Fig.3.
We find that the torus reconstructed by the interpolation technique 
corresponds well to the (true) torus derived directly by numerically following 
the orbit.
Furthermore we estimate the coefficients of the generating function at 
$J_1' \cong 0.11$ from the interpolation technique using the values at 
$J_1' \cong 0.04$ and $J_1' \cong 0.20$.
Using this generating function, we reconstruct the target torus 
for the loop-type orbit shown in Fig.3 by the square symbols on the loop-type torus. We find that the torus reconstructed 
using the interpolation technique corresponds well to the (true) torus 
for the loop-type orbit.

Let us summarize the main point that we get in the investigation of 
the interpolation method;
it is not necessary to calculate $S_{\bf n}$ at all values of
${\bf J'}$ by the procedure (i) $\sim $ (v) and we can get 
$S_{\bf n}$ for almost all values of J' by the interpolation technique. 
This fact reduces the computational time and amount 
of computer memory in making the generating functions.
We conclude that the torus-fitting method is a practicable method for 
obtaining the relations between the action variables and the 
Cartesian coordinates. 

\subsection{Application to Miyamoto-Nagai potential and strongly anisotropic potential}

In this subsection, we show the torus-fitting method is applicable to
other Galactic potential models and works very well.
First, we consider 
Miyamoto-Nagai potential given by,
\begin{equation}
\Phi = - \frac{1}{\sqrt{x^2+(a+\sqrt{y^2+b^2})^2}}, 
\end{equation}
where $a$ and $b$ are constants
(Miyamoto \& Nagai 1975; Binney \& Tremaine 1987).
We set $a=b=0.2$, and the surface of section in this model with total 
energy $E \cong -1.1405$ is shown in Fig.6.
The tori reconstructed by the torus-fitting method are shown by the 
plus symbols on the surface of section. The solid curves represent 
the (true) tori derived by the numerical calculations of the orbits. 
We find that the tori reconstructed by the torus-fitting method 
correspond well to the true ones both for box and loop orbits.

Furthermore we show in Fig.7 that the coefficients of the generating 
function for the box-type orbits are smooth functions of ${\bf J'}$. 
The five coefficients taken in descending order of the 
amplitudes of $S_{\bf n}$ are shown and they are calculated by the 
procedure (i) $\sim$ (v) at $J_1' \cong 0.22,0.33,0.43,0.53$.
The plus symbols represent the amplitudes of the coefficients at 
these values of ${\bf J'}$. Furthermore, in Fig.7, each solid line 
shows the linear interpolation line connected with the plus symbols 
for each coefficient. We can see that the interpolation lines are 
very smooth functions of ${\bf J'}$ and so the interpolation method 
for getting the values of the coefficient for any value of 
${\bf J'}$ can be used. 
We calculate the values of the coefficients derived by the procedure 
(i) $\sim$ (v) at $J_1' \cong 0.28, 0.38, 0.49$. 
These values are shown in Fig.7 by the crosses.
We can see that these marks correspond very well to the interpolation 
lines and so we confirm that the interpolation technique works very 
well also in the case of Miyamoto-Nagai potential.
We therefore confirm that the torus-fitting method works well
also in the Miyamoto-Nagai potential.
\begin{figure}
 \begin{center}
  \includegraphics[width=85mm]{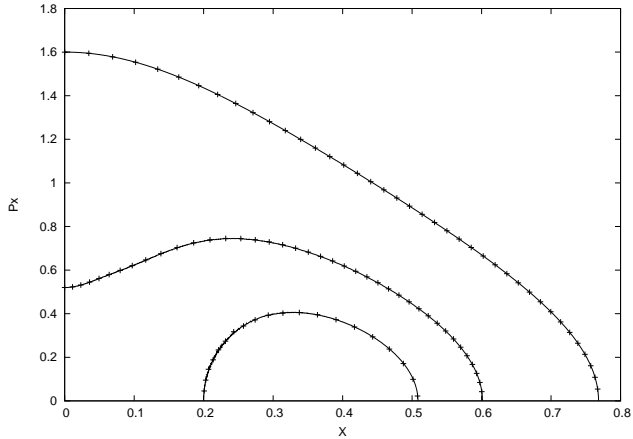}
 \end{center}
\caption{ The surface of section
with $y=0$ in the Miyamoto-Nagai 
potential with $E \cong -1.14$.
The solid curves represent the target tori derived by the numerical
calculations of the orbits of
three test particles. The plus symbols show the reconstructed 
target tori by use of the
torus-fitting method. 
The outermost curve corresponds to the tori at $J_1 \cong 0.54$, the middle
curve corresponds to the tori at $J_1' \cong 0.22$ and the innermost curve 
corresponds to the tori at $J_1' \cong 0.053$.
}
\end{figure}
 
\begin{figure}
 \begin{center}
  \includegraphics[width=60mm,angle=-90]{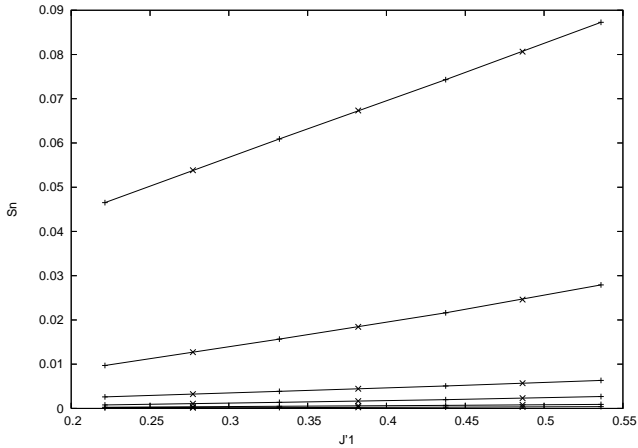}
 \end{center}
\caption{
The same figure as shown in Fig.4, but in Miyamoto-Nagai potential.
The amplitudes of $S_{\bf n}$ are calculated by the procedure 
(i) $\sim$ (v) at $J_1' \cong 0.22,0.33,0.43,0.53$.
The plus symbols represent the amplitudes of the coefficients at 
these values of 
${\bf J'}$. Each solid line shows the linear interpolation line 
connected with the plus symbols for each coefficient. 
Furthermore the values of the coefficients derived by the procedure 
(i) $\sim$ (v) at $J_1' \cong 0.28, 0.38, 0.49$ are shown 
by the crosses.
}
\end{figure}

Next we show that the torus-fitting method can be applied to the logarithmic
potential with low q-value, that is, asymmetric flat potential.
Here we set $q=0.4$ in the logarithmic potential with 
total energy $E \cong -0.337$ as a target Hamiltonian. 
The surfaces of section for this case are shown in Fig.8.
The solid curves represent the tori derived by the numerical 
calculations of the orbits
of the test particles.
The plus symbols show the tori constructed by the torus-fitting method.
Each reconstructed tori at each ${\bf J'}$ corresponds well to the 
true tori.
Hence we confirm that the torus-fitting method works well 
when the potential is asymmetric and flat.
However it should be noted that resonant orbits
besides major orbit families (box and loop orbits) appear in this case
although we omit the resonant tori on the surface of section shown in Fig.8.
In the next section, we show how the torus-fitting method can be applied to 
resonant tori and the method works well for the resonant orbits.

\begin{figure}
 \begin{center}
  \includegraphics[width=85mm]{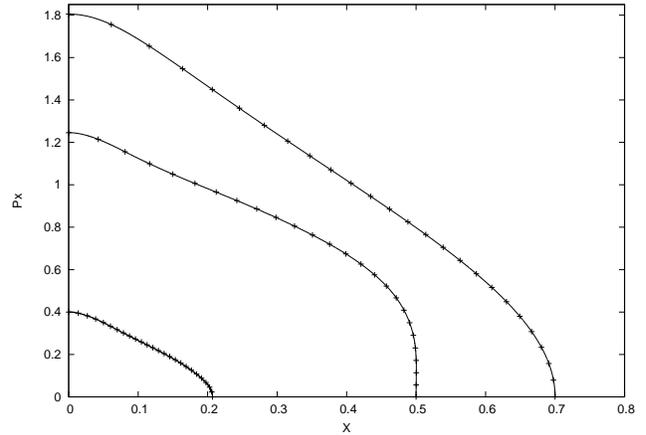}
 \end{center}
\caption{
The same surface of section 
as shown in Fig.1, but in the logarithmic potential with $q=0.4$.
The solid curves represent the target tori derived by the numerical
calculation of the orbits of
three test particles. The plus symbols show the reconstructed target 
tori by use of the
torus-fitting method. 
The outermost curve corresponds to the torus at $J_1' \cong 0.49$, the middle
curve corresponds to the torus at $J_1' \cong 0.28$ and the innermost curve 
corresponds to the torus at $J_1' \cong 0.032$.
}
\end{figure}

\section{Resonant Orbit}

\subsection{Formalism}

When the parameter $q$ in the logarithmic potential is sufficiently smaller than 1, there 
appear many resonant orbits clearly.
Fig.9 shows the surface of section for the 
logarithmic potential
with $q$ = 0.6, and
we can see two resonant tori\footnote{1:2 and 2:3 resonant tori} in this figure.
Here we explain procedures for how to 
construct a resonant torus using the torus-fitting method. The strategy is given as follows:

\begin{figure}
 \begin{center}
  \includegraphics[width=85mm]{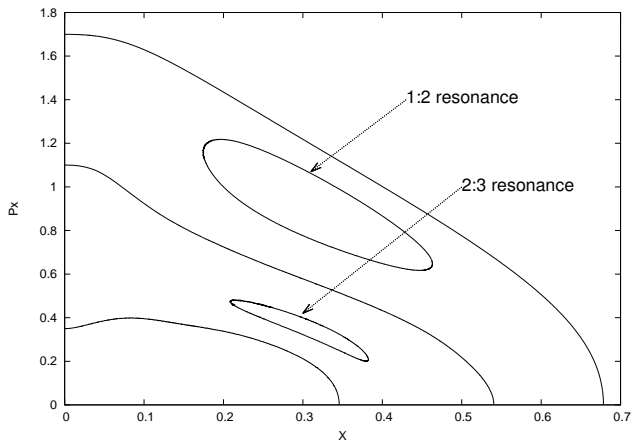}
 \end{center}
\caption{
The same surface of section 
as shown in Fig.1, but in the logarithmic potential with $q=0.6$.
The solid lines represent the target tori derived by the numerical
calculation of the orbits of test particles. 
Two ``islands'' represent resonant tori.
}
\end{figure}
To explain our strategy clearly, we here focus on  the 1:2 resonant torus, which is the largest island of the surface of section shown in Fig.9.
This resonant torus does not circle around the original point (0,0) on the surface of section unlike a box-type torus. That is, the angle variables that represent a position on the resonant torus do not cover a full range of the values of the angle variable ($0 \sim 2 \pi$), which is necessary to derive the Fourier coefficients $S_{\bf n}$ of equation (4). Furthermore, one value of the angle variable represents two points on the resonant torus. This means that in general, a position on the resonant torus is a two-valued function of the angle variable and so the position cannot be determined uniquely by one value of the angle variable. These two facts of the resonant torus make it impossible to determine the Fourier coefficients $S_{\bf n}$ of equation (4), so that we cannot apply directly the torus-fitting method to the resonant torus. 

To apply the torus-fitting method to the resonant torus, we introduce the following procedure.
First, we use an additional curve, which is a closed curve circled around the origin $(0,0)$. 
This additional curve is explained as follows: We draw two straight lines that pass through the origin and also contact 1:2 resonant torus that are shown as dashed line in Fig.10. The additional curve is determined so as to pass through the two points that divide the resonant torus into two parts.

 Second, we construct the two pseudo-tori from the resonant torus and the additional curve, which are shown in Fig.11, namely, one pseudo-torus consists of the additional curve (except the additional curve inside the resonant tours) and the upper part of the resonant torus (pseudo-torus 1), and another consists of the additional curve (except the additional curve inside the resonant torus) and the lower part of the resonant torus (pseudo-torus 2).  In our analysis, an elliptic curve is used as the additional curve, because we can analytically get the closed curve that passes the two contact points on the resonant torus. This elliptic curve used as the additional curve is shown in Fig.10 as the dotted curve. Finally  we obtain two pseudo-tori that are the closed curve whose shapes are similar to those to those of an ordinary box-type torus on the surface of section. Because these pseudo-tori have the full range of the values of the angle variable and the position on each pseudo-torus is a single valued function of the angle variable, we can get the Fourier coefficients $S_{\bf n}$ of equation (4).
  

\begin{figure}
 \begin{center}
  \includegraphics[width=85mm]{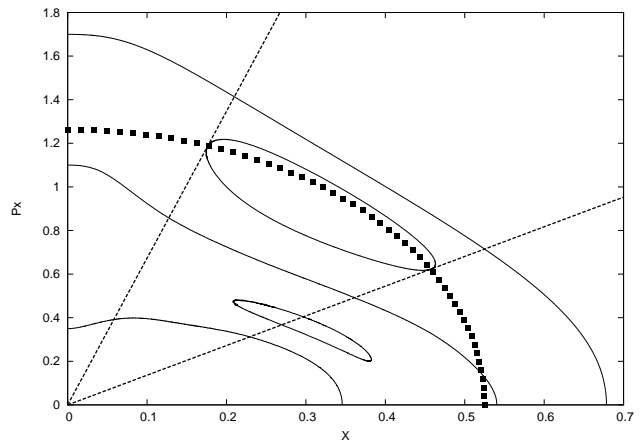}
 \end{center}
\caption{
The same surface of section 
as shown in Fig.9.
The solid lines represent the target tori derived by the numerical
calculation of the orbits of test particles. The dashed lines pass through the origin (0,0) and contact with the resonant torus. The dotted curve represents the additional curve that is used to reconstruct the resonant torus.
}
\end{figure}

 
We show the concrete way to reconstruct the resonant torus by the torus-fitting method with the use of the pseudo-tori mentioned above.
First we consider the pseudo-torus 1 and store
some phase space positions $({\bf x},{\bf p})$ on this torus.
Following the procedures (iii) $\sim$ (v) in the torus-fitting method, 
we can obtain the coefficients of the generating
function for the pseudo-torus 1. In this case, harmonic oscillator type is adopted as the toy 
Hamiltonian. Using these coefficients, the upper part of the pseudo torus 1 is reconstructed. We repeat the same procedure as mentioned just above and also obtain the coefficients of the generating function for the pseudo-torus 2. Finally the lower part of the pseudo torus 2 is reconstructed.

We find from the numerical calculations that
the pseudo-torus 1 has $J_{11}' \cong 0.34$, and the pseudo-torus 2 
has $J_{12}' \cong 0.3$. These values depend on the shape of the additional curve, but the difference between these values $J_{11}' - J_{12}'$ is independent of the shape of the additional curve.
By combining these two reconstructed pseudo-tori and cutting the 
part of the additional curve, we finally get the reconstructed resonant torus.
We can also reconstruct other types of resonant tori, e.g., 2:3 
resonant torus by the
same procedure as shown above, and the results are represented 
by the plus symbols in Fig.12. 
We find from Fig.12 that the reconstructed resonant tori correspond 
well to the true resonant tori and so we conclude that the torus-fitting method works well. 
 We focus on some typical resonant orbits in the resonant orbit family, and estimate coefficients of generating functions by using this method. As in the case of the major orbital families, we can obtain a family of the resonant torus (tori with the same type (e.g., 1:2) of the resonant torus) with interpolation technique for coefficients of generating functions, and reconstruct the family of the resonant orbit completely. 

\begin{figure}
 \begin{center}
  \includegraphics[width=80mm,angle=0]{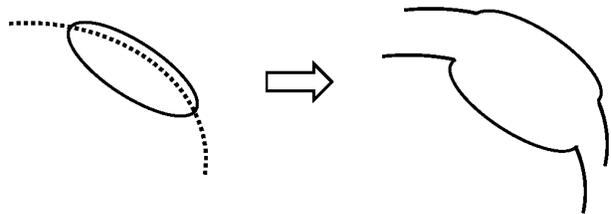}
 \end{center}
\caption{
Illustration of the resonant torus and the additional curve is shown 
at left.
The resonant torus can be reconstructed by the two pseudo-tori.
One of the pseudo-tori (pseudo-torus 1) consists of the additional curve (except for the additional curve inside  the resonant tours) and 
the upper part of the resonant torus as shown at upper right.
Another one (pseudo-torus 2) consists of the additional curve (except the additional curve inside the resonant tours) and the lower 
part of the resonant torus as shown at lower right. 
}
\end{figure}

\begin{figure}
 \begin{center}
  \includegraphics[width=85mm]{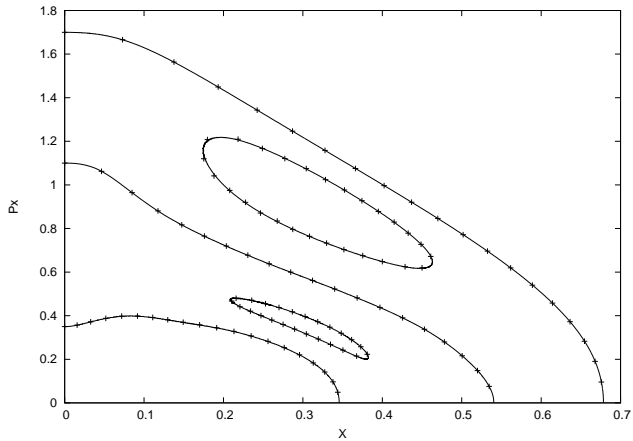}
 \end{center}
\caption{
The same surface of section 
as shown in Fig.10.
The solid lines represent the target tori derived by the numerical
calculation of the orbits of test particles. 
The plus symbols show the reconstructed target tori containing resonant tori reconstructed using the procedure 
mentioned in the text. See the text for details.
}
\end{figure}

Furthermore we confirmed that the torus-fitting method can reconstruct "small islands" representing 7:6 resonant tori that appear around the 1:2 resonant torus. 
We can see from Figure 13 that the higher resonant orbit appears around the 1:2 resonant torus, and the torus-fitting method can be applied to these "small islands" by the same procedures shown above.
Fig.14 represents one of the "small islands" in Fig.13, and plus symbols show the represent reconstructed small island using the torus-fitting method. Therefore, we conclude that the torus-fitting method is still useful to reconstruct minute structures.

\begin{figure}
 \begin{center}
  \includegraphics[width=85mm]{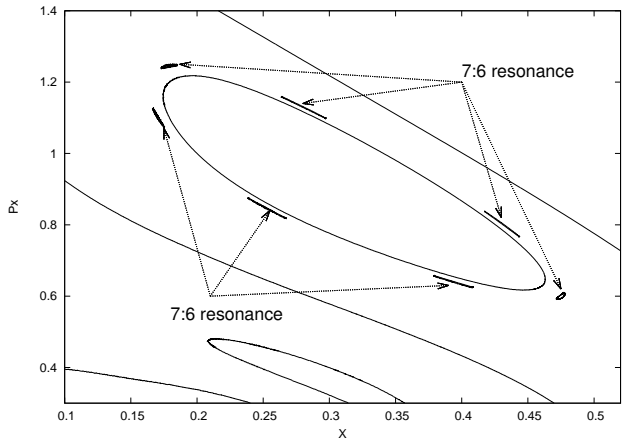}
 \end{center}
\caption{
The same surface of section 
as shown in Fig.9.
The solid curves represent the target tori derived by the numerical
calculations of the orbits of test particles. 
The ``small islands'' representing small resonant tori appear around 1:2 resonant torus. 
}
\end{figure}

\begin{figure}
 \begin{center}
  \includegraphics[width=85mm]{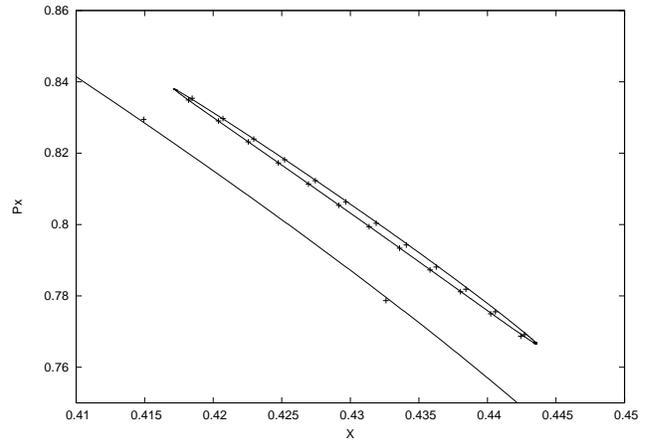}
 \end{center}
\caption{
The same surface of section 
as shown in Fig.13, but magnified around one of "small islands".
The solid curves represent the target tori derived by the numerical
calculation of the orbits of test particles. 
The plus symbols show the reconstructed target tori by using the
torus-fitting method. 
}
\end{figure}

\subsection{Some comments on the torus-fitting method}

Before leaving this section, two subtle points concerning the torus-fitting method is considered.
We first mention the application of the torus-fitting method to more complicated structures.
Phase-space structures are in general complicated fractal structures under a general gravitational potential (Binney \& Tremaine 1987).
As shown in Fig.14, the torus-fitting algorithm is, in principal, applicable to such complicated phase-space 
structures, and this is one of the advantage of this method, while we need more CPU time for numerical calculations to get very fine structures in our method. However, we need not reproduce very fine structures on the phase-space in applying the fitting method to construct Galactic models that should be compared with observational data. This is, because, very fine structures cannot be reconstructed by the smearing effect due to observational errors. Hence, in the practical use of the fitting method, it is sufficient to construct torus structures whose scales on the phase-space are larger than those of the fine structures smeared by observational errors. So, in practical applications of constructing the torus structures of a Galactic model, it is enough to consider major orbits (box and loop orbits) and lower resonant orbits whose sizes of the tori are enough large to be considered.

We next mention angular variables. Although obtaining relations 
between angular variables and Cartesian coordinates is necessary to understand 
dynamical features of torus, this is irrelevant to the main subject. 
This is because that the purpose of this paper does not reproduce all characters 
of an invariant torus, but for obtaining relations between ${\bf J}$ and 
$({\bf x},{\bf p})$ through generating functions. 
In particular, the fact that we can reproduce any tori if some typical generating 
functions that reproduce some representative tori is important, and on 
this account we do not treat the angular variables. 
However, since winding number is an important quantity that represents 
dynamical features of an invariant torus, this is worth a mention in passing. 
In general, the winding number is a quantity that characterizes torus 
structure. For example, 
if a winding number is rational, the corresponding torus becomes a point or 
a set of points on a two-dimensional Poincare section, and we do not take 
account of these structures, because weights of these become zero when we construct a 
distribution function. On the other hand, if the winding number is irrational, 
the corresponding torus becomes a one-dimensional curve on a two-dimensional 
Poincare section. Because it is necessary to treat this case, the winding 
number is important when we use the torus-fitting method. As the purpose 
of this paper is to obtain the relation between
${\bf J}$ and $({\bf x},{\bf p})$, it is not necessary to 
show up the winding number of any torus, and we may leave the details 
to this topic.

\section{Summary and Discussion}

In this paper, we propose a new method, that is, the torus-fitting 
method for obtaining generating functions in two-dimensional 
Galactic potentials.
We confirmed that the torus-fitting method works very well for 
constructing tori of major families of orbits in the two-dimensional 
logarithmic potential and Miyamoto-Nagai potential.
In this method, the coefficients of generating 
functions are smooth functions of action variables ${\bf J'}$ if the type of torus
is the same type. Hence we can obtain coefficients of generating functions at 
any value of ${\bf J'}$ by interpolating coefficients calculated at 
some typical values of ${\bf J'}$. This fact reduces the computational time and the amount 
of memory required for computers.
So this method is more practical compared with the direct numerical 
calculations of $ J_i = \frac{1}{2 \pi} \oint_{\gamma_i} {p} d {q}$ 
for obtaining the relations between the action variables of the target 
Hamiltonians and the Cartesian coordinates.
Furthermore, the torus-fitting method is still applicable to 
resonant orbit families besides major orbit families, although we use 
the additional technique in which we use the pseudo-tori for constructing 
the target resonant tori. 
Hence the torus-fitting method is useful for analyzing 
a real Galactic system in which a lot of resonant orbits exist.

Here, we discuss applications of the torus-fitting method to observational data. To understand the dynamical structures of the Galaxy, we first assume a theoretical dynamical model, that is, a gravitational potential of the Galaxy, which should be compared with observations. If it is necessary for us to get the types of the orbits of each observed star and its value of ${\bf J'}$ in the assumed model, how can we get them without direct numerical integrations of the orbits while the observations can provide only the positions and velocities of the stars at a given time? The strategy is given as follows: First, we suppose that the type of orbit of all observed stars is the box type as a trial. The use of the torus-fitting method makes it possible to convert the Cartesian coordinates, $( {\bf x},{\bf p} )$, of the observed stars into the action variables, ${\bf J'}$. In this way, we can get the values of ${\bf J'}$ of the observed stars if all orbits were box-type orbits. On the other hand, we have already estimated the allowed region of the values of ${\bf J'}$ for the box-type orbit in the process of the construction of the generating functions when we apply the torus-fitting method to this assumed model. So if the estimated value of ${\bf J'}$ of an observed star is included in this allowed region for the box type, we can recognize that the type of star is the box type and this value of ${\bf J'}$ is the true value of the action variable of the star. Otherwise, the assumption that this starfs orbit is the box type is not true. So we again compute the value of ${\bf J'}$ under the other supposition that the star's orbit is a loop-type orbit or a resonant orbit by trial and error. In this way, we can finally derive the true value of ${\bf J'}$ and the types of orbits.

If we assume that the Galaxy has a steady state, and almost all orbits of the celestial objects (the stars and dark matter) in the Galaxy are regular, then, as described in $\S 1 $, the phase-space distribution function of the objects in the Galaxy is a function of three independent isolating integrals, which correspond to action variables. So we theoretically construct phase-space distribution functions of the action variables for any Galactic model depicts all orbits as regular. This means we need to recognize the values of the action variables of the observed stars when we compare a theoretically constructed phase-space distribution function with the distribution of the observed stars. Hence it is necessary and important to convert the Cartesian coordinate, $({\bf x},{\bf p})$, of the observed stars into the action variables, ${\bf J'}$. As mentioned above, we can do so by the torus-fitting method.

It is apparent that the torus-fitting method cannot be applied to systems in which 
chaotic orbits are dominant in the phase space.
However, if almost all orbits move for long periods around their nearby tori although 
the orbits are strictly chaotic, the orbits can be regarded as being approximately 
regular ones.This may be the case for some galactic bulges and some kinds of 
elliptical galaxies. The reason is that some galactic bulges and some 
elliptical galaxies have anisotropic velocity dispersions that cause 
the triaxial shape of the structures. This suggests that these systems 
have approximately three isolating integrals. That is, almost all orbits 
can be regarded as being approximately regular ones.
If this guess is true, the torus-fitting method can be applied 
to these systems.

We finally discuss future work on the torus-fitting method.
As a first step, we examined the two-dimensional potentials in this paper, although, a real galaxy generally has a three-dimensional potential.
So we will try to apply the torus-fitting method to some three-dimensional 
potentials and a forthcoming paper will present
this application. Modern space astrometry projects will provide us reliable
information about stellar phase-space coordinates in the Galaxy,
and so the torus-fitting method is useful for examining
steady-state dynamical models of the Galaxy.

\section*{Acknowledgments}

This was supported by the JSPS KAKENHI Grant Number 23244034(Grant-in Aid  for Scientific Research (A)),

\bsp

\label{lastpage}

\end{document}